\documentclass[a4paper,fleqn,usenatbib]{mnras}
\usepackage{lscape,graphicx}
\usepackage{rotating}
\usepackage{color}
\usepackage{threeparttable}
\usepackage{amsmath,amssymb}

\usepackage{color}

\def\apj{ApJ}
\def\apjs{ApJS}
\def\mnras{MNRAS}

\def\apjl{ApJ}
\def\aap{A\&A}
\def\aj{AJ}

\def\araa{ARA\&A}

\def\asec{\ifmmode ^{\prime\prime}\else$^{\prime\prime}$\fi}

\def\it{\sl}
\def\degs{\ifmmode ^{\circ}\else$^{\circ}$\fi}
\def\amin{\ifmmode ^{\prime}\else$^{\prime}$\fi}
\def\asec{\ifmmode ^{\prime\prime}\else$^{\prime\prime}$\fi}
\def\fss{\hbox{$.\!\!^{\rm s}$}}        
\def\fdg{\hbox{$.\!\!^\circ$}}          
\def\farcs{\hbox{$.\!\!^{\prime\prime}$}}  
\def\h{$^{\rm h}$}
\def\m{$^{\rm m}$}

\def\degs{\ifmmode ^{\circ}\else$^{\circ}$\fi}
\def\amin{\ifmmode ^{\prime}\else$^{\prime}$\fi}

\def\eqalign#1{\null\,\vcenter{\openup1\jot \m@th
   \ialign{\strut\hfil$\displaystyle{##}$&$\displaystyle{{}##}$\hfil
   \crcr#1\crcr}}\,}
\def\j1932{J1932}

\title[The $\gamma$-ray pulsar J1932+1916 in X-rays]{Observations of the $\gamma$-ray pulsar J1932+1916 in X-rays}
\author[A.~Karpova, P.~Shternin, D.~Zyuzin, A.~Danilenko, Yu.~Shibanov]{
A.~Karpova$^{1}$\thanks{E-mail: annakarpova1989@gmail.com},
P.~Shternin$^{1}$, 
D.~Zyuzin$^{1}$, 
A.~Danilenko$^{1}$, 
Yu.~Shibanov$^{1,2}$\\
$^{1}$Ioffe Institute, Politekhnicheskaya 26, St. Petersburg, 194021, Russia\\
$^{2}$Peter the Great St. Petersburg Polytechnic University, Politekhnicheskaya 29, St. Petersburg, 195251, Russia}

\date{Accepted XXX. Received YYY; in original form ZZZ}
\pubyear{2016}

\begin{document}

\label{firstpage}
\pagerange{\pageref{firstpage}--\pageref{lastpage}} 
\maketitle

\begin{abstract}   
We present the analysis of the archival \textit{Suzaku} and \textit{Swift} X-ray observations of the
young $\gamma$-ray pulsar J1932+1916 field. 
The data revealed a point-like object at the $\gamma$-ray position of the pulsar
and diffuse X-ray emission around it.
Spectra of the point-like source and diffuse emission are well-described by absorbed  power-law models  
with spectral parameters typical for pulsar plus pulsar wind nebula systems. 
Therefore we suggest that \textit{Suzaku} and \textit{Swift} detected the X-ray counterpart of
PSR~J1932+1916. 
Assuming this interpretation, we constrain the distance to the pulsar in the range of 2--6~kpc. 
We also suggest possible association of the pulsar with the nearby supernova remnant
G54.4$-$0.3 and discuss its implications for the pulsar proper motion, age and distance.

\end{abstract}


\section{Introduction}

The $\gamma$-ray pulsar PSR J1932$+$1916 (hereafter \j1932) was recently discovered by
\citet{Pletsch2013} in the \textit{Fermi} Large Area Telescope data 
using a novel blind search algorithm initially dedicated for  gravitational
wave  astronomy \citep{Pletsch2012}. 
Utilising  the cloud computing with the Einstein@Home\footnote{http://einstein.phys.uwm.edu}
volunteer-based supercomputing system,  
\citet{Pletsch2013} found four new young and energetic $\gamma$-ray pulsars. 
\j1932\ is the brightest and youngest of them. 
Subsequent radio observations failed to detect the pulsar \citep{Pletsch2013}.
\textit{Fermi} timing analysis yields the pulsar position
R.A.=19\h32\m19\fss70(4) and Dec.=+19\degs16\amin39\asec(1)\footnote{Numbers in parentheses are 1$\sigma$ 
uncertainties corresponding to last significant digits quoted.} ($l=54\fdg 7$ and
$b=0\fdg 08$).
The pulsar has the characteristic age $\tau = 35$~kyr, the period 
$P=208$~ms, the spin-down luminosity $\dot{E}=4\times10^{35}$~erg~s$^{-1}$ and 
the magnetic field $B=4.5\times10^{12}$~G. 
The so-called ``pseudo-distance''
estimate based on the empirical  correlation between $\gamma$-ray fluxes and
distances of pulsars \citep{abdo2013ApJS} results in the distance $D\sim 1.5$~kpc.   

Relatively small distances and ages usually found for
\textit{Fermi} pulsars make these objects promising  for X-ray studies. 
In the \textit{Swift} X-ray Telescope (XRT) Point Source
Catalogue \citep{evans2014},  
we found a source 1SXPS J193219.4+191635 
which position agrees  with the \j1932\ \textit{Fermi} position 
allowing us to consider it as the pulsar X-ray counterpart candidate. 
Since there are only 9 source counts detected with XRT, one cannot make any definite conclusions on source properties.
Fortunately, the \j1932\ field was also observed in 2010 with \textit{Suzaku} as a part
of  follow-up  X-ray observations of unidentified \textit{Fermi} sources. 

Here we report the analysis of the \textit{Swift} and \textit{Suzaku}    
observations. The
observations and data reduction are described in
Section~\ref{S:observations}. The \textit{Suzaku} image creation procedure 
is presented in Section~\ref{S:imaging}. We found an extended emission 
at the \j1932\ position in  \textit{Suzaku} images. 
We argue  that the \textit{Swift} point source is likely the
pulsar \j1932\ counterpart, while the extended emission detected with
\textit{Suzaku} is the associated pulsar wind nebula (PWN). 
Its spatial structure and spectral properties
are analysed in Sections~\ref{S:extent} and \ref{S:XraySpectra}. 
We discuss our results in Section~\ref{S:Discussions} and summarise them in
Section~\ref{S:Conclusions}.

\section{Observations and data reduction}\label{S:observations}
\subsection{\textit{Swift} data}\label{S:SwiftObs}
\begin{figure}
\begin{minipage}[h]{1.0\linewidth}
\center{\includegraphics[width=0.9\linewidth, trim = {2.7cm 0 0 0.8cm},clip]{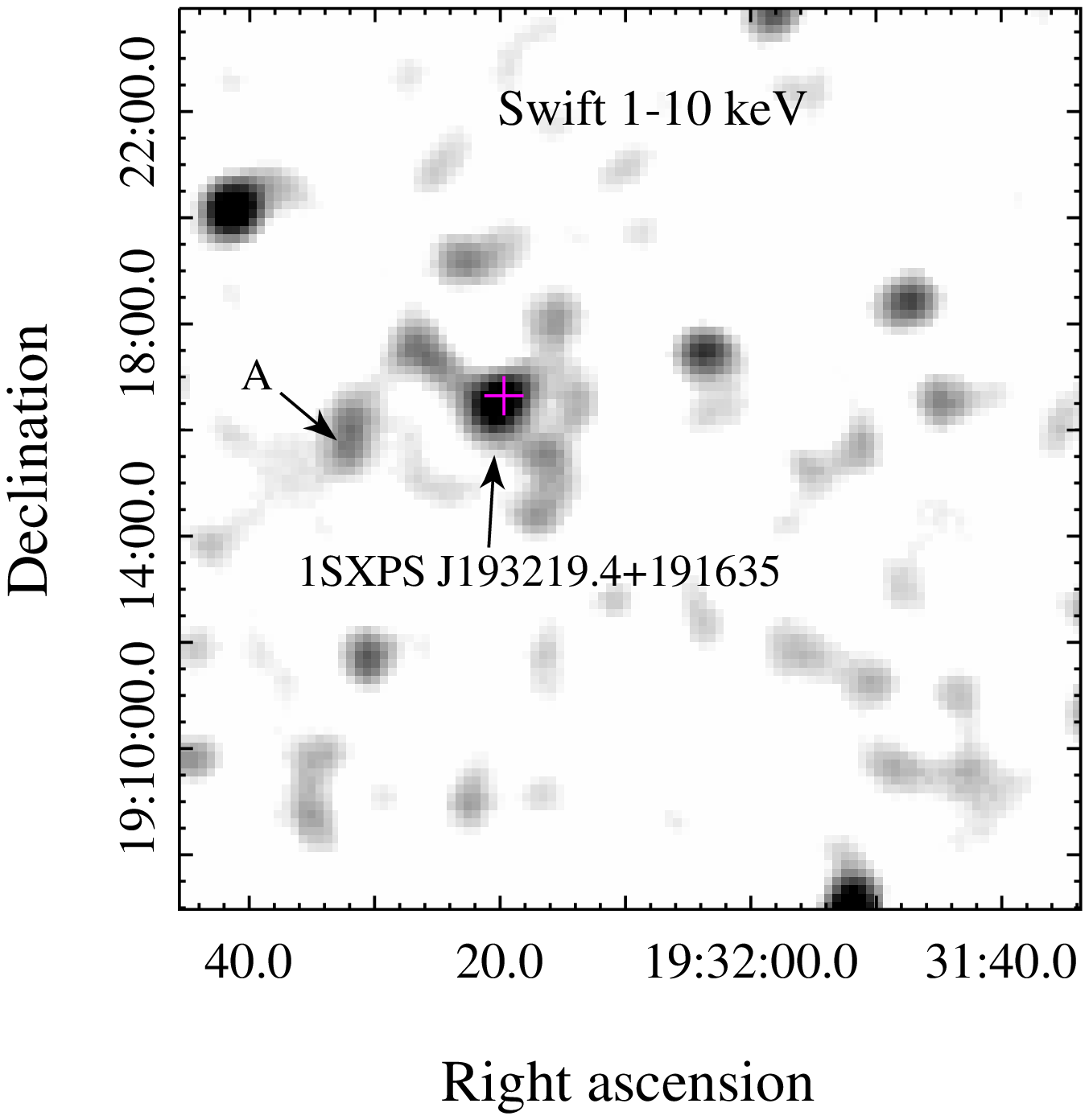}}
\end{minipage}
\begin{minipage}[h]{1.0\linewidth}
\center{\includegraphics[width=1.05\linewidth,trim = {0 0 0 3.7cm},clip]{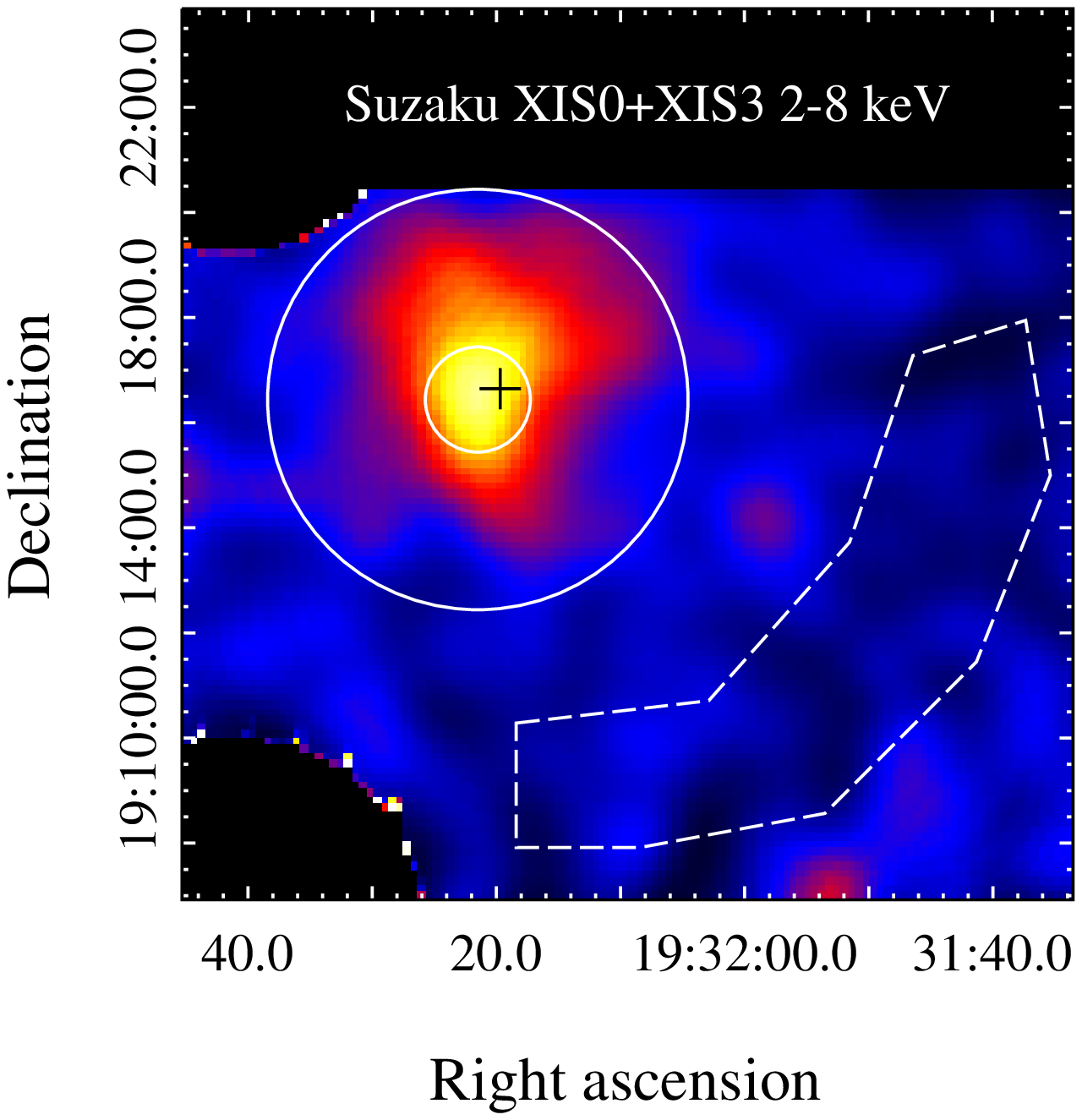}}
\end{minipage}
\begin{minipage}[h]{1.0\linewidth}
\center{\includegraphics[width=1.05\linewidth,trim = {0 0 0 3.7cm},clip]{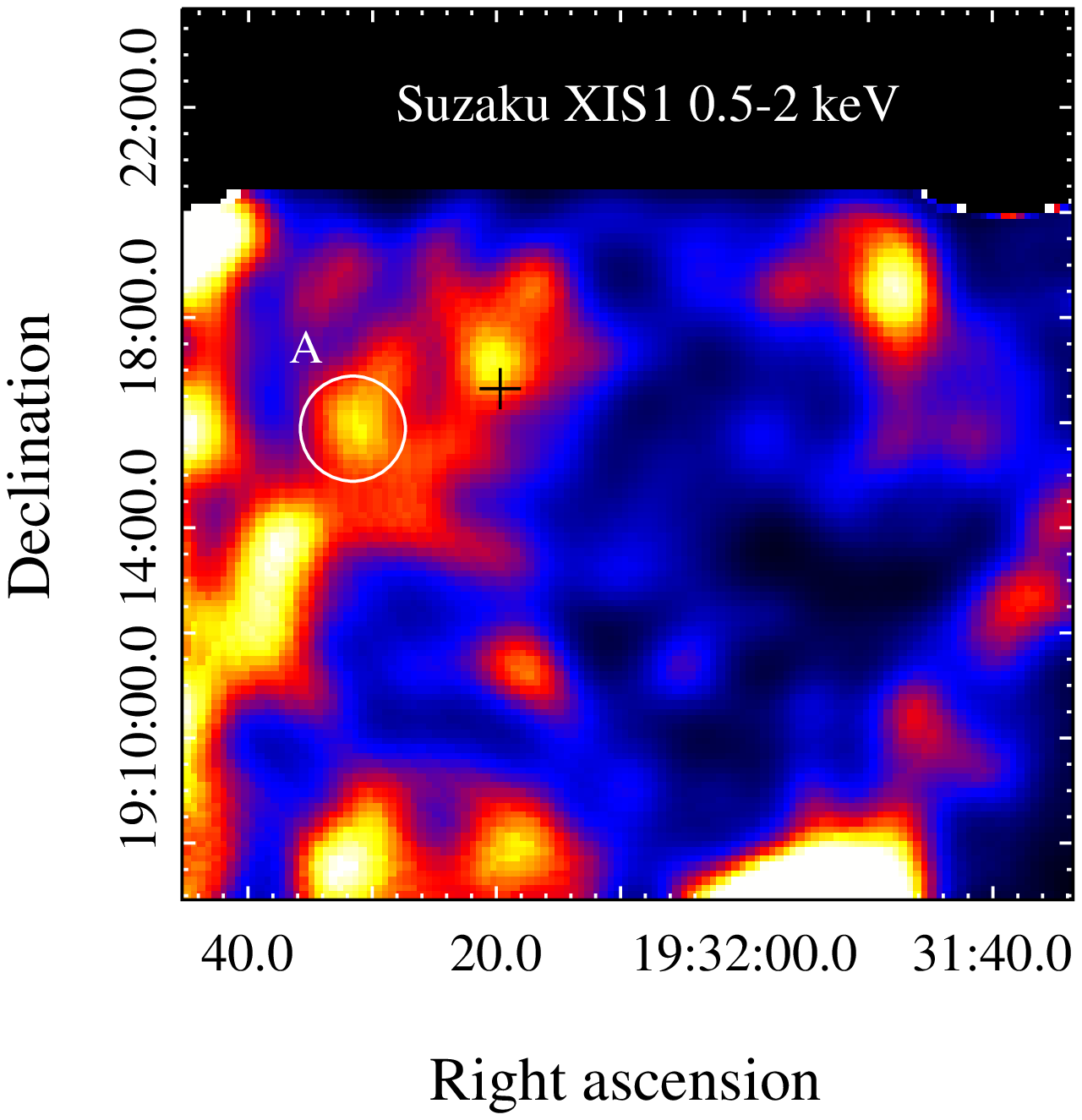}}
\end{minipage}
\caption{The PSR J1932$+$1916 field as observed with
\textit{Swift} XRT in 1--10~keV (top). The image is binned by 4 pixels and
smoothed with a 5-pixel Gaussian kernel. 
The 1SXPS J193219.4+191635 source is marked.  
The 2--8 keV {\it Suzaku} XIS0+XIS3 (middle)
and 0.5--2 keV XIS1 (bottom) images of the PSR J1932$+$1916 field. The
images are binned by 8 pixels and smoothed with a Gaussian
function of $\sigma=4$ pixels. Linear brightness scale is used for
the top and middle images and squared brightness scale is used for the
bottom image. The {\it Fermi} position of
PSR~J1932$+$1916 \citep{Pletsch2013} is shown by the crosses. In the
middle panel, regions used for spectral analysis are shown
(see text for details). The soft source `A' falling in the extraction region 
is marked in the top and bottom panels. 
}
\label{fig:images}
\end{figure}

The \j1932\ field was observed with \textit{Swift} three times. 
The possible X-ray counterpart 1SXPS
J193219.4+191635 is detected only in the longest December 2010  observations
(ObsID 00041803001, exposure $\approx 10$~ks), still the detection significance is poor with the
mean source count rate of $(1.3\pm 0.6)\times 10^{-3}$~cts~s$^{-1}$.
Other observations are too shallow ($<0.6$~ks) and were ignored.
We retrieved the reduced XRT data for the  ObsID 00041803001 from the \textit{Swift} archive.
The respective image is shown in the top panel of Figure~\ref{fig:images} where 
the source 1SXPS J193219.4+191635 is marked. Its astrometrically-corrected 
coordinates \citep[][Section 3.7]{evans2014} provided by the catalogue  
are R.A.=19\h32\m19\fss58 and  Dec.=+19\degs16\amin30\farcs4 with the position error circle radius of 
7.2 arcsec (90 per cent confidence). The \j1932\ \textit{Fermi} position lies within 
this error circle and is marked by the cross. 

\subsection{\textit{Suzaku} data}\label{S:SuzakuObs}


The 45~ks \textit{Suzaku} observations\footnote{ObsID 405028010, PI T.
Yasuyuki.} of the \j1932\ field were carried out on 2010, April 27
and were pointed at R.A.=293\fdg04 and Dec.=19\fdg26.  The
\textit{Suzaku} X-ray Telescope consisted of three
front-illuminated X-ray Imaging Spectrometers (XIS0, XIS2 and
XIS3) and one back-illuminated (XIS1), one of which, namely XIS2,
was not operating since 2006 due to a damage event. In addition,
a part of XIS0 chip was unavailable due to
similar event occurred in
2009.\footnote{http://www.astro.isas.ac.jp/suzaku/doc/suzakumemo/\\ \indent suzakumemo-2010-01.pdf}

For our analysis, we used the {\sc heasoft} v.6.17 software. We
reprocessed the unfiltered data using {\sc aepipeline} tool. We
applied the standard screening criteria, i.e. we filtered flickering and
hot pixels, time intervals corresponding to \textit{Suzaku}
passage of the South Atlantic Anomaly, the Earth elevation angles
$<$5\degs\ and the Earth day-time elevation angles $<$20\degs. 
Data of editing modes $3\times 3$ and $5\times 5$ were combined. After the procession,   
the effective exposure decreased down to about 24 ks. 
The decrease in exposure time is mainly due to the  elevation angle constraints 
and the requirement of the instantaneous pointing to be within 1.5 arcmin of the mean.

\section{Data Analysis}

\subsection{\textit{Suzaku} Imaging}\label{S:imaging}

The images combined from two front-illuminated detectors XIS0 and
XIS3 are shown in the middle panel of Figure~\ref{fig:images}
for the 2$-$8 keV band. The image in the 0.5--2 keV band from the
back-illuminated detector XIS1 that provided higher sensitivity at
lower energies is shown in the bottom panel of
Figure~\ref{fig:images}. Constructing these images, we
subtracted the non-X-ray background (NXB) generated via the
{\sc xisnxbgen} task \citep{Tawa2008}. The resulting images were
binned by 8 arcsec per pixel and smoothed with a Gaussian function
of $\sigma=4$ pixels width. 
The vignetting correction was then applied following the approach
similar to one used by \citet{fujinaga13} and \citet{izawa15}.
This procedure requires a flat-field image construction on the
basis of the appropriate astrophysical background (ABKG) spectral model.

For all XIS detectors, we extracted the  ABKG spectra 
from the off-source polygon-shaped region shown in the
middle panel of Figure~\ref{fig:images}. The detector
responses were generated by the task {\sc xisrmfgen} and the
Ancillary Response Files (ARFs) were
generated by the task {\sc xissimarfgen} assuming a uniform sky
source. Then the NXB-subtracted ABKG spectra from XIS~0--3 detectors were 
fitted simultaneously with {\sc xspec} v.12.9.0 \citep{Arnaud1996}. 
The employed spectral model consists of the Cosmic X-ray Background (CXB) and
thermal plasma components. It is believed that CXB is isotropic
but its contribution for a particular line of sight depends on the total Galactic
absorption for this direction. Thus, we fixed this component
in the fit assuming the absorbed power-law spectrum with the photon index
$\Gamma=1.4$, the surface brightness $I=5.4\times
10^{-15}$~erg~s$^{-1}$~cm$^{-2}$~arcmin$^{-2}$ \citep[in $2-10$~keV
band;][]{Ueda1999} and the hydrogen column
density $N_{\rm H}=2.8\times
10^{22}$ cm$^{-2}$. The latter value represents the total Galactic absorption in the \j1932\ direction 
and is discussed in
detail in Section~\ref{S:dist}. The remaining part of the spectrum
was fitted with the absorbed three-temperature optically thin
thermal plasma model 3-T {\sc apec}. 
The best-fit parameters
are given in Table~\ref{t:best-fit}. A lower number of components
(e.g., 2-T model) resulted in
statistically unacceptable fits. The primarily use of this model is to
obtain a reliable approximation of the observed astrophysical
background for the flat-field image construction and we 
do not justify its physical nature here.  
Finally, we simulated the
flat field image with {\sc xissim} tool \citep{Ishisaki2007}
using the best-fit spectral model for the background. The photon
number in simulations was large enough to ensure the good
statistics. This approach allowed us to correctly account for the
vignetting in the extended source area.

\subsection{Source extent}\label{S:extent}

\begin{figure}
\begin{minipage}[h]{1.0\linewidth}
\center{\includegraphics[width=1.07\linewidth,clip]{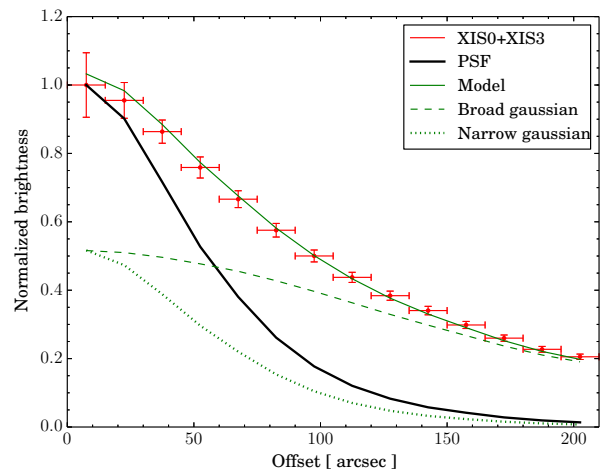}}
\end{minipage}
\caption{Comparison of the observed radial brightness 
profile (crosses) of the \textit{Suzaku} \j1932\ counterpart candidate   
with the simulated PSF (black solid line) and the best-fit 
spatial model (green solid line). Narrow and broad Gaussian model components
are shown with dotted and dashed lines, respectively, see text for details.}
\label{fig:profiles}
\end{figure}

The resulting images demonstrate the presence of an X-ray excess
spatially coincident with the \textit{Fermi} position of \j1932
(lower panels of Figure~\ref{fig:images}). However, the excess can
not be described by the emission from  a point source. This is illustrated in
Figure~\ref{fig:profiles} where we compare the radial brightness
profile in the XIS0+XIS3 image extracted from the concentric
annuli centred at the source brightness peak (crosses) with the radial profile of the
\textit{Suzaku} point spread function (PSF) shown with the black
solid  line. The latter is created with the
\textsf{xissim} tool. It is clear that the X-ray emission extent
is significantly larger than the PSF half-power diameter of 2 arcmin.
Note that both NXB and ABKG  are subtracted
in Figure~\ref{fig:profiles}.

A relatively low spatial resolution of \textit{Suzaku} does not allow to perform
detailed morphological analysis of extended sources. 
However, an approximate spatial model of the source can be
inferred. The model is useful to estimate the source extent and is
necessary to obtain ARFs 
for the spectral analysis. For these purposes, we constructed
NXB-subtracted images from XIS0 and XIS3 data in the 2--8 keV
band. This band was chosen to eliminate the contribution from a soft point-like
source `A' clearly seen in the \textit{Swift} and XIS1 images of Figure~\ref{fig:images}. 
This source is listed as 1SXPS J193231.7+191540 in the 1SXPS catalogue.
The ABKG images generated as discussed in the previous section
were subtracted. The images were then binned and smoothed in the same way as before.
The resulting 
images were fitted with the
{\sc ciao} v.4.7 {\sc sherpa} software using simulated PSF images as
convolution kernels. We found that the spatial brightness distribution of the considered source 
can be described with the  model consisting of two 2D-Gaussian functions with negligible ellipticities:  
a broad one with the full width at half maximum (FWHM)~$\approx$~4.5 arcmin and a narrow one with FWHM~$\lesssim $~0.5 arcmin. 
Their centres are  shifted by about 70 arcsec relative to each other.  
In Figure~\ref{fig:profiles}, the radial profiles of the best-fit model and 
its two components convolved with the PSF 
are shown with green solid, dashed and dotted lines, respectively. 
The broad emission component can be interpreted as the PWN    
powered by \j1932.  
The narrow component peak position offsets by about 30 arcsec  from the point-like \textit{Swift} counterpart candidate position.  
The offset significance is only about 1.5$\sigma$ taking into account
the 90 per cent XIS positional uncertainties of 19 arcsec  \citep{uchiyama2008} 
and  the 7.2 arcsec \textit{Swift} position error. 
It is thus natural
to assume that  the narrow component represents the pulsar itself. 
On the other hand, it may  just be a bright compact feature of the  PWN. 
Better spatial resolution observations are necessary to verify that.

\subsection{X-ray spectral analysis}\label{S:XraySpectra}

We assume that  the X-ray emission around the \j1932\ position
has contributions from the point-like pulsar presumably detected with 
\textit{Swift}  and  from the extended source seen with
\textit{Suzaku}. In what follows, we call the point-like component
as `PSR' and the extended component as `PWN'. 
According to the analysis in Section~\ref{S:extent},
the PSR component must contribute to the central part of the 
extended source in \textit{Suzaku} data. Bearing this in mind, we divided the extended source region
in two parts shown in the bottom-left panel of
Figure~\ref{fig:images}: the central 60 arcsec-radius circle and the
ring with radii of 60 arcsec and 4 arcmin. In the latter region, 
the 60 arcsec-radius circle containing the point-like source `A' was 
excluded from the spectral analysis. The \textit{Suzaku} spectra were
extracted from these two regions from all XIS detectors. NXB
spectra appropriate for each region   were
generated with {\sc xisnxbgen} for each detector and subtracted.

We simultaneously fitted all 9 (3 regions $\times$ 3 XIS
detectors) \textit{Suzaku} spectra assuming the PSR, PWN and ABKG
contributions in the central circle and in the
ring and the ABKG contribution in the polygon region.
Using {\sc xissimarfgen} tool, for each detector we constructed seven ARFs
for each contribution.
For the PWN component, we generated ARFs adopting the broad
Gaussian spatial model with $\sigma=$4.5 arcmin as derived in the
previous section. The narrow Gaussian was not included
since we attributed it to the pulsar.  
On the other hand, as noted in Section~\ref{S:extent}, it may represent the PWN compact structure.
However, we checked that the addition of this narrow Gaussian 
in ARFs generation for the PWN does not affect
final results significantly. For the PSR component, 
we generated point-source ARFs. Finally, for all regions, we generated ARFs for an
uniformly extended source to be used with the ABKG component. 

In order to put constrains on the point-source
contribution which follow from the \textit{Swift} observations, we
also included \textit{Swift} data in the spectral analysis. We
extracted spectrum from the
30 arcsec-radius aperture centred at the 1SXPS
J193219.4+191635 position 
and background was extracted from the source-free region in the
\textit{Swift} image.

The ABKG model was already described above in Sect \ref{S:imaging}. For the
PSR and the PWN components, we used the absorbed power-law (PL) models
with shared values of the hydrogen absorption column density.
To account for the photoelectric absorption we used the 
{\sc phabs} model with default cross-sections {\sc bcmc}
\citep{balucinska1992} and abundances {\sc angr}  \citep{anders1989}.
The parameters for the PSR model were tied between the
\textit{Swift} and  \textit{Suzaku} spectra. Different
contributions from the PWN to different
\textit{Suzaku} spectra are automatically taken into account by
the ARFs. In principle, some contribution from the PWN emission to
the \textit{Swift} source spectrum  can be expected. Basing on the spatial model of
Section~\ref{S:extent}, we found that $\approx 4$ per cent of the total PWN flux contribute to the 
\textit{Swift} extraction aperture. This was included in the spectral analysis. 
Using {\sc xrtmkarf} tool, we generated two XRT ARFs: for the extended emission to
account for the possible PWN contribution and for the point source PSR.

We used the unbinned spectra to fit for the model parameters. The 
{\sc xspec} software was applied to calculate the likelihood statistics.
$C$-statistics \citep{cash1979} was used for the
\textit{Swift} spectra, while the 
{\sc pgstat}\footnote{https://heasarc.gsfc.nasa.gov/xanadu/xspec/manual/\\ \indent XSappendixStatistics.html}
statistics was used for the \textit{Suzaku} spectra. The latter statistics corresponds
to a Poisson source with a Gaussian background and is appropriate
since the {\sc xisnxbgen} task returns the simulated NXB count rate
with Gaussian error-bars. 
Spectra were fitted in the 0.5--10~keV range. 
At higher energies, the \textit{Suzaku} spectra are dominated by NXB.
The estimation of the parameter uncertainties was performed via the Bayesian inference framework. 
We sampled the posterior parameter space using the Markov chain Monte-Carlo ({\sc mcmc}) procedure. Specifically,
we used the affine-invariant {\sc mcmc} sampler \citep{goodman2010CAMCS}
employed as a convenient python package {\sc emcee} by
\citet{foreman2013}.
Best-fit parameters and the 90 per cent
uncertainties are presented in Table~\ref{t:best-fit}. 
The latter were estimated as the 90 per cent highest posterior density intervals of 
the corresponding marginal posterior density distributions. 

\begin{figure}
\begin{minipage}[h]{1.0\linewidth}
\center{\includegraphics[width=0.71\linewidth,clip,angle=-90, clip]{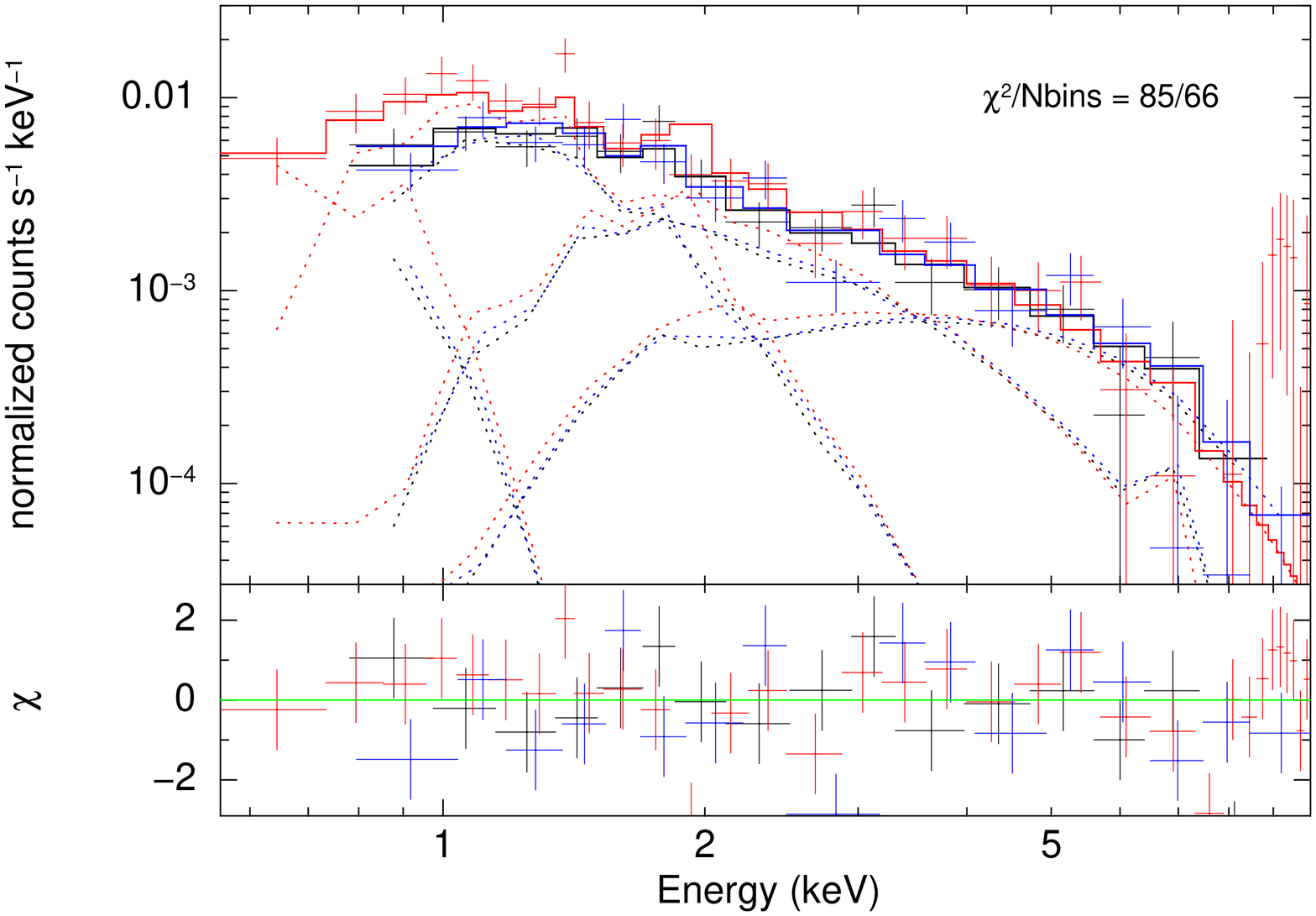}}
\end{minipage}
\begin{minipage}[h]{1.0\linewidth}
\center{\includegraphics[width=0.71\linewidth,clip,angle=-90,clip]{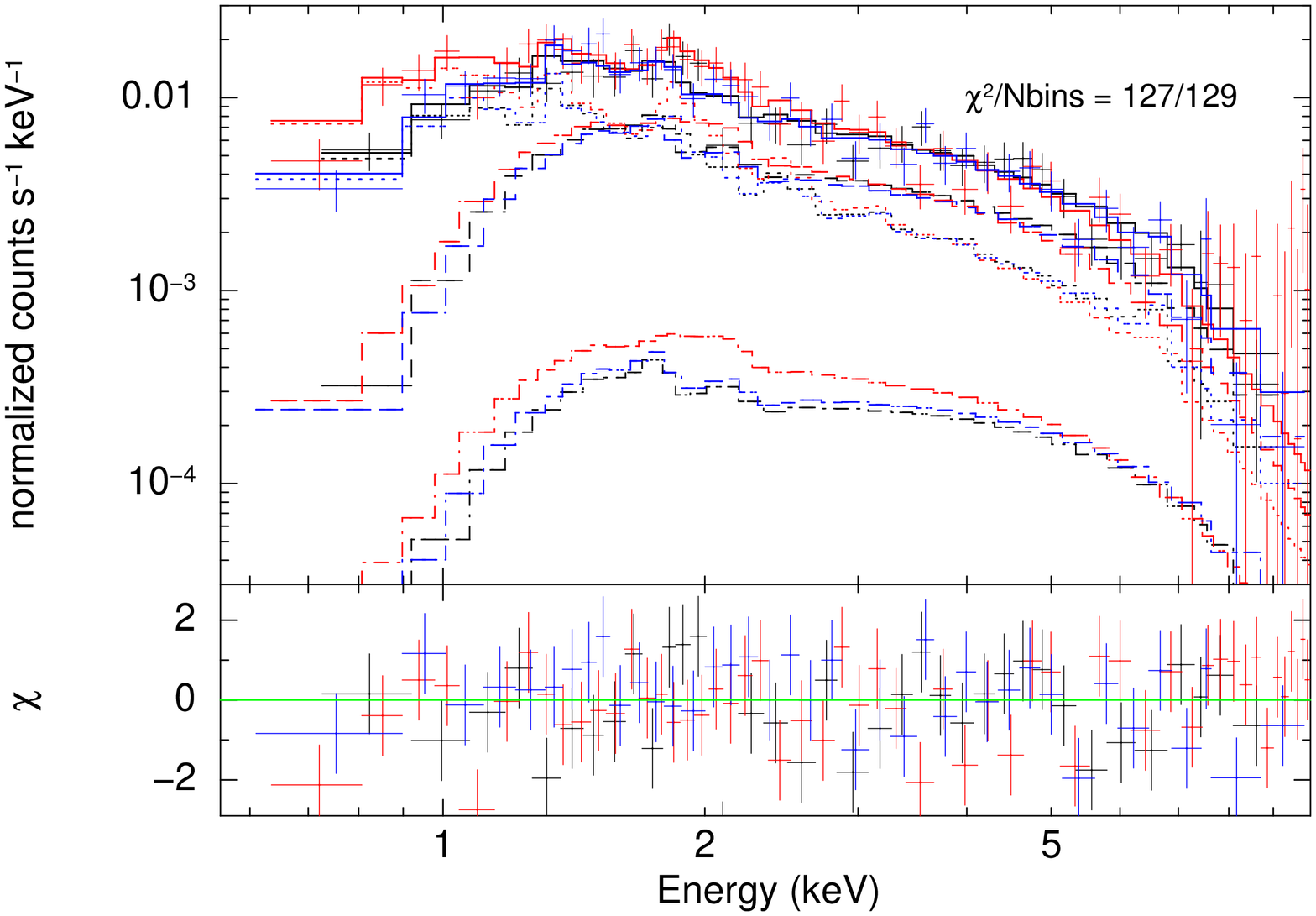}}
\end{minipage}
\begin{minipage}[h]{1.0\linewidth}
\center{\includegraphics[width=0.71\linewidth,clip,angle=-90,clip]{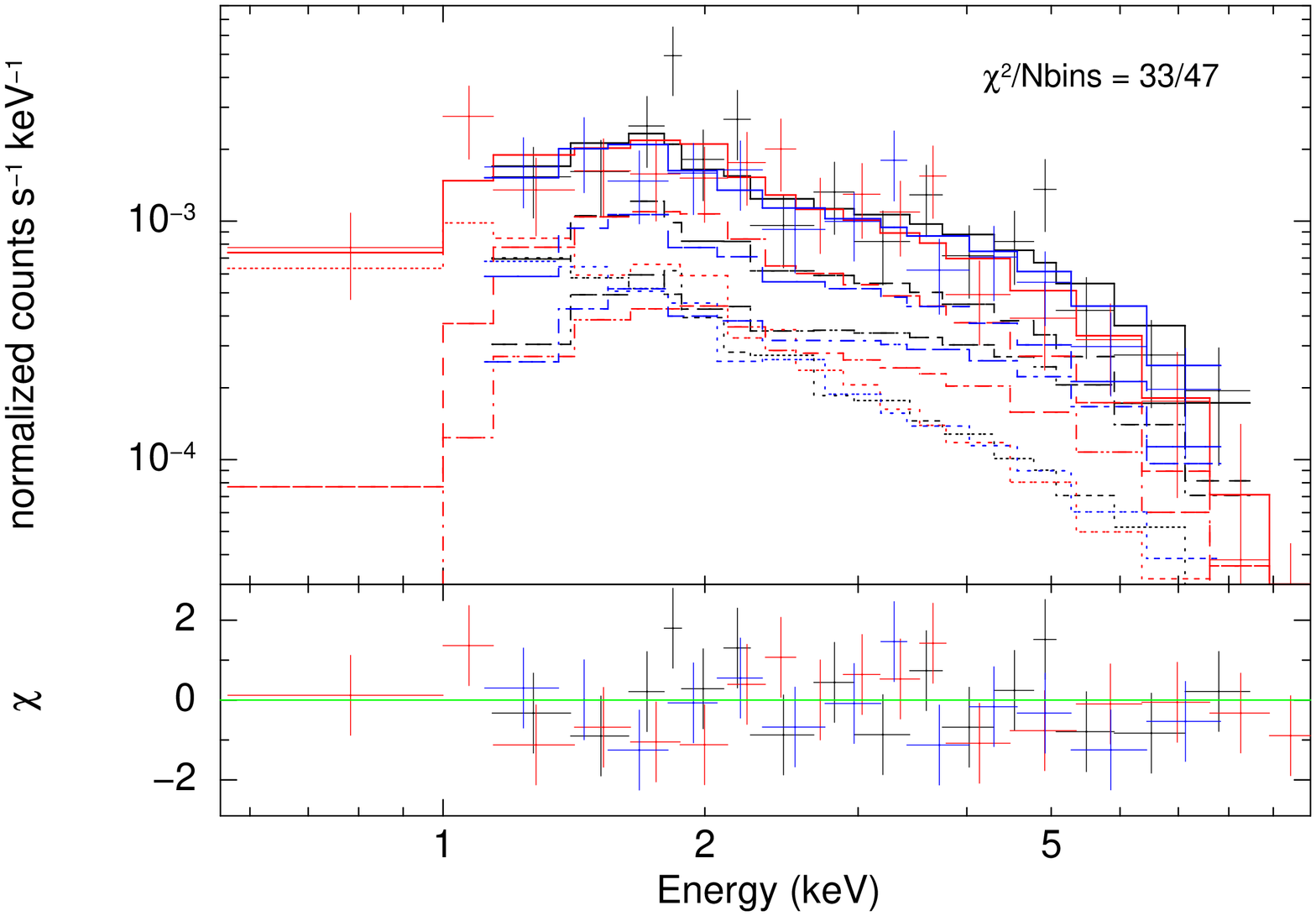}}
\end{minipage}
\caption{ Observed spectra (crosses) and best fit models (solid lines) 
extracted from the polygon (top panel), the ring
(middle panel) and the central circle (bottom panel) regions 
shown in Figure~\ref{fig:images}. 
XIS0, XIS1 and XIS3 data are presented  
by black, red and blue colours, respectively.
Low sub-panels demonstrate fit residuals
normalised by the data uncertainties in each spectral bin ($\chi$). In the top and middle panels, the
spectra are binned for illustrative purposes 
by 30  counts per bin, while in the bottom panels --  by 10 counts per bin.    
Dotted lines
in the top panel represent four ABKG  model components. 
In the middle and bottom panels, dotted, dashed and dot-dashed lines represent 
the ABKG, the PWN and the PSR contributions, respectively.
The total $\chi^2$ values and the numbers of energy bins (Nbins) 
are shown in top sub-panels for each spectral extraction region.}

\label{fig:spectra}
\end{figure}

The goodness of the fit is illustrated in Figure~\ref{fig:spectra} where the spectra 
from the polygon  (top panel), 
the ring (middle panel) and the central circle (bottom panel) regions are compared with the best-fit spectral models. 
For illustration purposes, the polygon and circle spectra are binned to ensure 30 counts in each energy bin, 
and the ring region spectra are binned to ensure 10 counts per energy bin. 
We also show the corresponding $\chi^2$ values in each plot. As seen from Figure~\ref{fig:spectra}, 
the model describes the observed data reasonably well. We do not see any significant unmodelled features in the residuals. 
The excess in the residuals at energies $>$7 keV for the XIS1 spectra is due to incomplete NXB modelling 
at high energies \citep[see, e.g.,][]{izawa15}. 
We also assessed the goodness of fit for our model by calculating the 
posterior predictive p-value \cite[e.g.,][]{GelmanBook} for the the 
Pearson $\chi^2$ test statistics ({\sc pchi} in {\sc xspec}) appropriate for the Poisson data. 
We used {\sc xspec fakeit} tool to generate N=256 synthetic spectra for the model parameters 
sampled from the posterior distributions. 
We found that only in 68 per cent of simulations, 
the test statistics value for the synthetic data was smaller than for the observed data. 
This means that our model cannot be rejected.

\begin{table}
\caption{Best-fit parameters for the ABKG, the presumed PWN and the PSR.
All errors are at 90 per cent confidence.}
\label{t:best-fit}
\begin{center}
\begin{tabular}{ll}
\hline
\multicolumn{2}{c}{ABKG} \\
\hline
\multicolumn{2}{l}{} \\
$N_{\rm H_1}^a$, 10$^{22}$ cm$^{-2}$ & $0.8^{+0.3}_{-0.3}$\\
$T_1^b$, keV                         & $0.06^{+0.02}_{-0.01}$\\
$N_{1}^c$, 100 cm$^{-5}$              & $0.6^{+38}_{-0.5}$\\
$N_{\rm H_2}$, 10$^{22}$ cm$^{-2}$ & $1.0^{+0.2}_{-0.2}$ \\
$T_2^b$, keV                       & $0.5^{+0.1}_{-0.2}$\\
$N_{2}^c$, 10$^{-2}$ cm$^{-5}$     & $2^{+2}_{-1}$\\
$N_{\rm H_3}$, 10$^{22}$ cm$^{-2}$  & $1.8^{+1.0}_{-0.7}$ \\
$T_3^b$, keV                       & $1.9^{+1.1}_{-0.5}$\\
$N_{3}^c$, 10$^{-3}$ cm$^{-5}$     & $11^{+4}_{-5}$\\
\multicolumn{2}{l}{} \\
\hline
\multicolumn{2}{c}{PWN} \\
\hline
\multicolumn{2}{l}{} \\
$N_{\rm H}^a$, 10$^{22}$ cm$^{-2}$                          & $1.2^{+0.3}_{-0.3}$\\
$\Gamma^d$                                                  & $1.8^{+0.4}_{-0.3}$\\
$N_{PL}^e$, 10$^{-4}$ photons s$^{-1}$ cm$^{-2}$ keV$^{-1}$ & $2.6^{+1.5}_{-1.0}$\\
$f_{X}^f$, $10^{-12}$ erg s$^{-1}$ cm$^{-2}$                & $1.2^{+0.3}_{-0.2}$\\
\multicolumn{2}{l}{} \\
\hline
\multicolumn{2}{c}{PSR} \\
\hline
\multicolumn{2}{l}{} \\
$\Gamma^d$                                                  & $1.4^{+1.0}_{-1.0}$\\
$N_{PL}^e$, 10$^{-5}$ photons s$^{-1}$ cm$^{-2}$ keV$^{-1}$ & $1.7^{+4.6}_{-1.3}$\\
$f_{X}^f$, $10^{-13}$ erg s$^{-1}$ cm$^{-2}$                & $1.3^{+0.9}_{-0.5}$\\
\end{tabular}
\end{center}
\begin{tablenotes}
\item $^a$ Hydrogen column density; for the PSR it is the same
as for the PWN
\item $^b$ Plasma temperature
\item $^c$ {\sc apec} normalization
\item $^d$ Photon index
\item $^e$ PL normalization
\item $^f$ Unabsorbed flux in the 0.5--8 keV band
\end{tablenotes}
\end{table}

\section{Discussion}\label{S:Discussions}

We found that the spectrum of the extended source detected with \textit{Suzaku} at the
$\gamma$-ray position of  the pulsar \j1932\ is well described by an absorbed
power law model with $\Gamma\approx 2$. This value is typical
for  PWNe \citep{kargaltsev2008}. $N_\mathrm{H}$ value obtained from the spectral fits is twice lower 
than the total Galactic $N_\mathrm{H}$ in this direction excluding an extragalactic interpretation.
The pulsar itself is not 
directly identified with \textit{Suzaku} but is likely visible  with \textit{Swift} 
as a point-like source. 
Its spectrum can be
described by the absorbed  power-law, although the PL parameters are poorly constrained, i.e.,  
$\Gamma\approx 0.5-2.5$. Nevertheless,  the values obtained 
are not unreasonable neither for  pulsars nor for  compact PWNe. It is
instructive to transform the obtained X-ray fluxes into luminosities and to compare them
with the data on other young PSR+PWN systems. To do this, the reliable distance
estimate is required. 

\subsection{Distance to \j1932}\label{S:dist}
\begin{figure}
\begin{minipage}[h]{1.0\linewidth}
\center{\includegraphics[width=1.0\linewidth,clip]{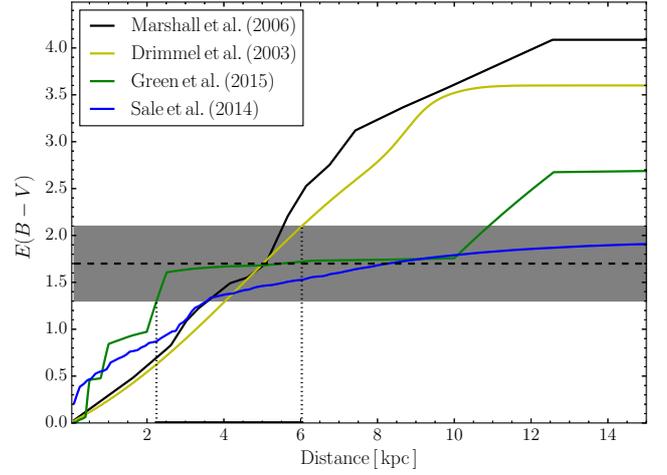}}
\end{minipage}
\caption{Extinction--distance relations in the \j1932\ direction according to various extinction maps.
The grey-filled horizontal strip corresponds to the \j1932\ extinction value 
inferred from the X-ray spectral fit. The adopted distance range is between the vertical dotted lines, 
see text for details.}
\label{fig:extmaps}
\end{figure}

Due to the absence of \j1932\ in the radio, 
the only available distance estimate $D\sim 1.5$~kpc was obtained  by \citet{Pletsch2013} using the empirical 
correlation between the $\gamma$-ray flux and the distance, although this
estimate is known to be very crude \citep[e.g.,][]{abdo2013ApJS}.
Assuming 100 per cent efficiency in
$\gamma$-rays, \citet{Pletsch2013} estimated upper limit of
$6.6$~kpc, neglecting, however, possible beaming of the emission.
If the $\gamma$-ray emission is highly beamed, the distance
 can be even higher.

Another possible way to estimate the distance is based on the
analysis of the interstellar extinction towards the source.
Existing 3D extinction maps of the Galaxy mainly trace the dust
absorption quantified by the selective extinction value $E(B-V)$.
The latter can be estimated using an empirical correlation between
$E(B-V)$ and $N_{\rm H}$. Here we use the relation $N_{\rm
H}=(0.7\pm 0.1)\times E(B-V)\times 10^{22}$~cm$^{-2}$ given by
\citet{Watson2011} basing on the analysis of the gamma-ray bursts.
This relation is consistent with one given by \citet{guver2009}
constructed using the optical and X-ray observations of supernova
remnants (SNRs).\footnote{We rely on the work of
\citet{guver2009} instead of their more recent work \citet{foight2016} since
\citet{guver2009} use the same abundances from
\citet{anders1989} as we do, while \citet{foight2016} use abundances from \citet*{wilms2000}.}
With this relation in hand, the total Galactic $N_{\rm H}$  can be
readily obtained from the total $E(B-V)$ value. 
The total selective extinction in the \j1932\ direction
according to the Galactic Dust Reddening and Extinction 
Service\footnote{https://irsa.ipac.caltech.edu/applications/DUST/} 
is $E(B-V)\approx 4$ \citep{schlafly2011}, transforming to the total $N_{\rm H}\approx
2.8\times 10^{22}$~cm$^{-2}$. We thus use this value in the CXB
spectral model in Sections~\ref{S:imaging}--\ref{S:XraySpectra}.

Our X-ray spectral fits yield $N_{\rm H}=(1.2\pm0.3)\times
10^{22}$~cm$^{-2}$ (Table~\ref{t:best-fit}) which transforms 
to $E(B-V)=1.7\pm0.4$ (1$\sigma$ errors). In Figure~\ref{fig:extmaps}, we compare 
this range (grey-filled strip) with several extinction-distance relations for the \j1932\ direction constructed 
using the {\sc python} package {\sc mwdust} \citep{Bovy2016}. 
The plot includes the extinction maps of \citet[][based on the reg-giant stars photometry in the 
2MASS survey]{Marshall2006}, \citet[][based on the main sequence star photometry in the Pan-STARRS and 
the 2MASS surveys]{dustmap2015},
\citet[][based on the INT/WFC Photometric H$\alpha$ Survey]{Sale2014} and 
\citet*[][based on an analytic model fit to the COBE DIRBE data]{drimmel2003AsAp}. The maps of \citet{Sale2014} 
and \citet{dustmap2015} are not applicable at distances $\gtrsim 5$~kpc in this direction due to paucity 
of the main-sequence stars in the surveys. On the other hand, \citet{Marshall2006} map is based on brighter 
reg giant stars and reliably traces the extinctions to larger distances. According to Figure~\ref{fig:extmaps}, 
it also agrees well with the map of \citet{drimmel2003AsAp}. However, at smaller distances ($\lesssim 2$~kpc), the map of
\citet{dustmap2015} is preferable for the similar reason: there are much more main sequence stars than red giants.

Therefore, we base our conservative lower limit of the distance estimate on the \citet{dustmap2015} map and 
the upper limit on the \citet{drimmel2003AsAp} map. 
This results in the \j1932\ distance range of $2-6$~kpc (bounded by vertical dotted lines in Figure~\ref{fig:extmaps}).
For convenience, below we use the notation $D_{4\rm kpc}\equiv D/(4~\mathrm{kpc})$.

\subsection{Luminosity, efficiency and PWN size }\label{S:lum}
\begin{figure*}
\begin{minipage}[h]{0.49\textwidth}
\includegraphics[width=1.1\linewidth, clip]{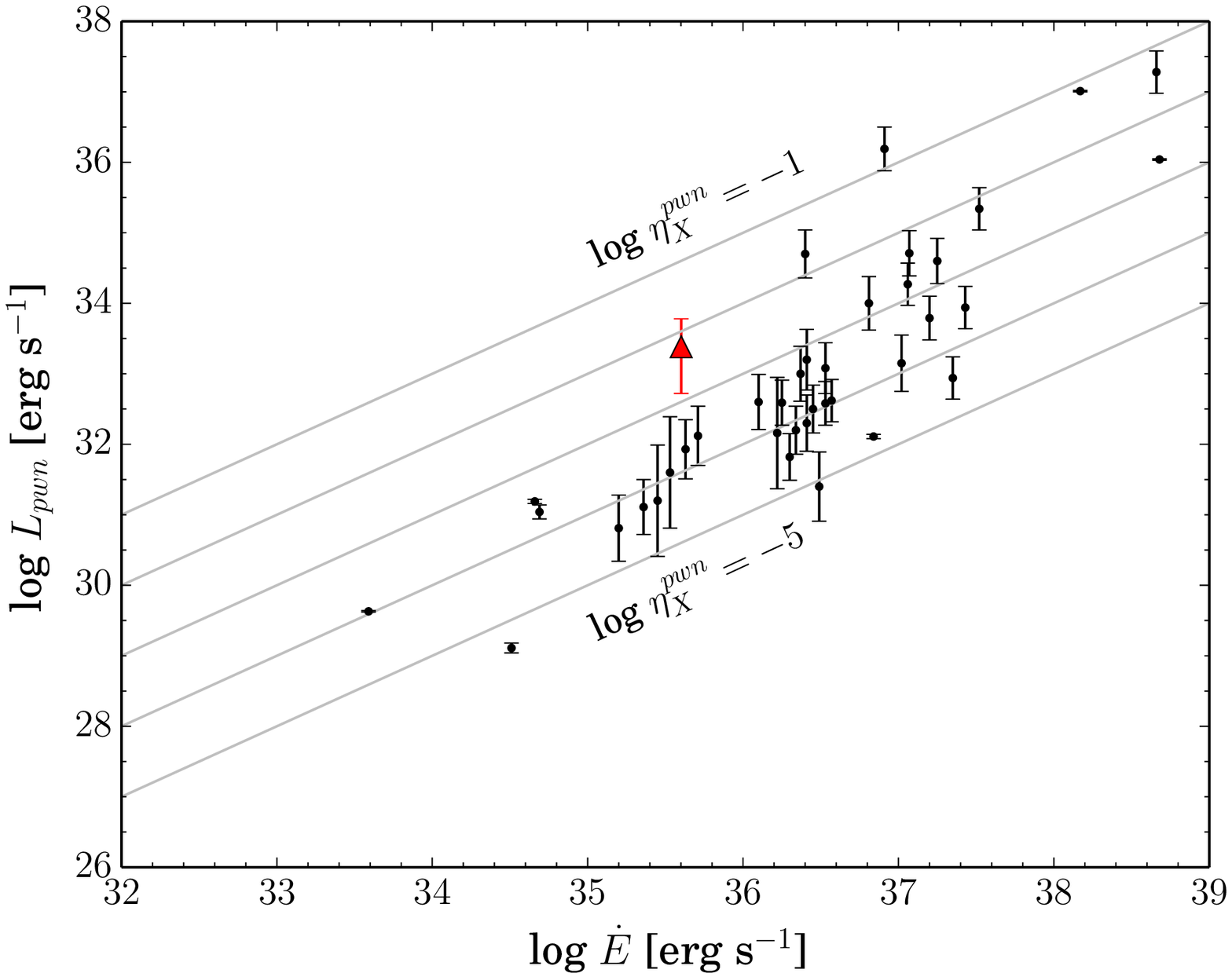}
\end{minipage}
\begin{minipage}[h]{0.49\textwidth}
\includegraphics[width=1.05\linewidth, clip]{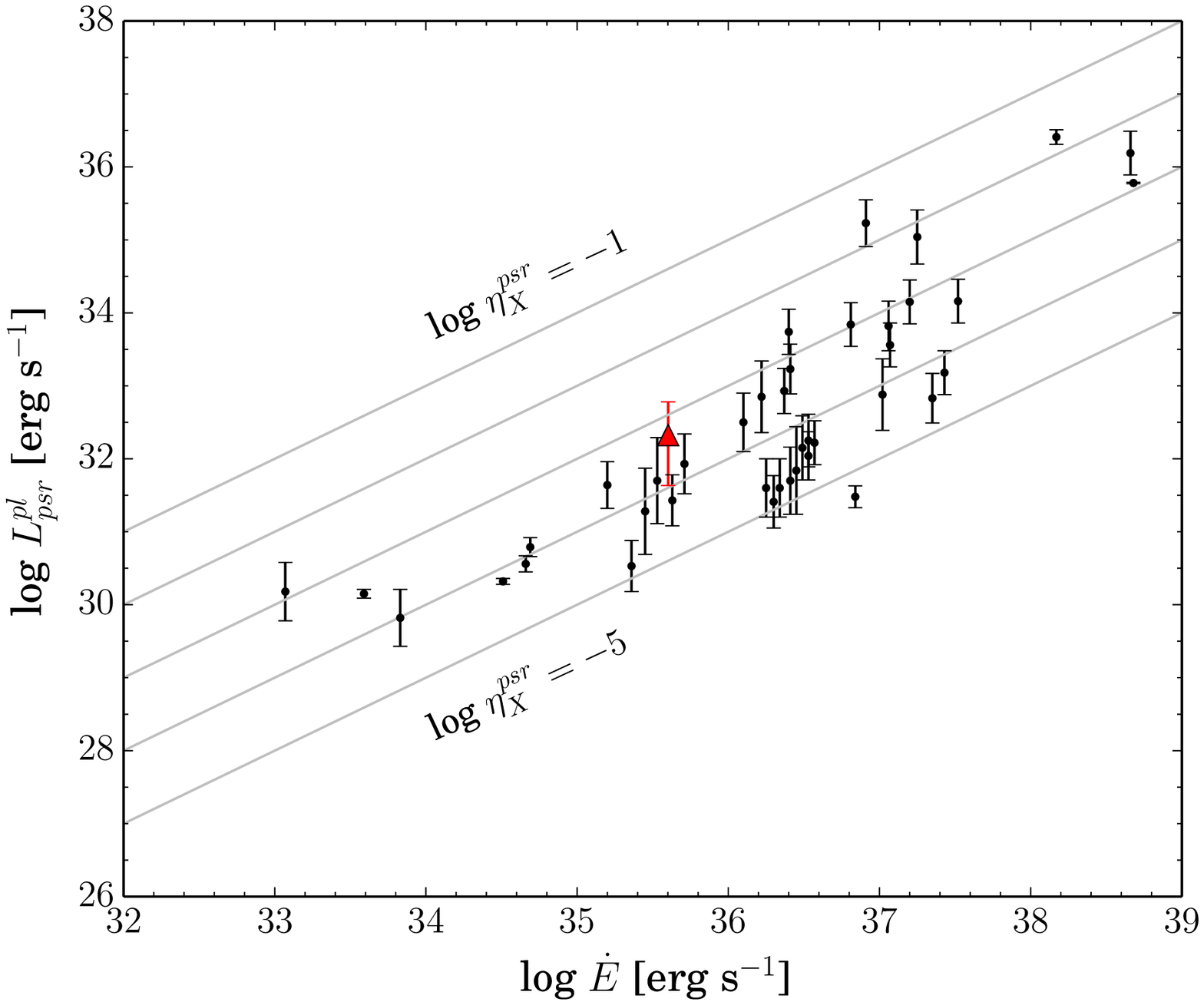}
\end{minipage}
\caption{PWN luminosities (\textit{left}) and PSR luminosities
(\textit{right}) versus spin-down energies for various objects. The
\j1932\ luminosities are shown by triangles for $D=4$~kpc. Error
bars combine the 1$\sigma$ flux uncertainties and the distance range
uncertainty. } 
\label{fig:LEdot}
\end{figure*}

Estimated X-ray luminosities of \j1932\ and 
its PWN are $L_{\rm PSR} =
2.5^{+1.7}_{-1.0}\times{10}^{32} D_{4\rm kpc}$~erg~s$^{-1}$ and  $L_{\rm PWN} =
2.3^{+0.6}_{-0.4}\times{10}^{33} D_{4\rm kpc}$~erg~s$^{-1}$.  
To compare them with the luminosities of other pulsar+PWN systems,
we used the data from \citet{kargaltsev2008} with updates from \citet{Shibanov2016}.
The plots of the  pulsar and PWN luminosities versus the spin-down luminosity 
are presented in Figure~\ref{fig:LEdot} where  \j1932\ 
is shown by the red triangle.
As seen, if the detected X-ray sources are indeed  related to \j1932, 
this pulsar  produces a brighter PWN in comparison to 
other ones with similar $\dot{E}$. 
However, this system is generally 
compatible with the overall $L-\dot{E}$ pulsar and PWN distributions.
The  ratio $L_{\rm PWN}/L_{\rm PSR} \approx 10$  for \j1932\ is 
also consistent with the  range $0.1\lesssim\L_{\rm
PWN}/\L_{\rm PSR}\lesssim10$ obtained by \citet{kargaltsev2008} using the current 
sample of these systems.   
This supports the pulsar and PWN  identification of the detected sources. 

For \j1932, the $\gamma$-ray flux is about $8\times10^{-11}$~erg~s$^{-1}$~cm$^{-1}$ in the 0.1--100 GeV range \citep{Pletsch2013}. Thus, the pulsar $\gamma$-ray to X-ray flux ratio is $\sim 600$ that is typical for $\gamma$-ray pulsars with similar $\dot{E}$ \citep{abdo2013ApJS}. 

\citet{bamba2010} pointed on a correlation between PWNe X-ray physical
sizes and pulsar characteristic ages. Accounting for 
the distance uncertainties, the \j1932\ PWN size \citep[defined as three times the Gaussian $\sigma$,][]{bamba2010} is 
$6.5\pm3.5$~pc. For the \j1932\ characteristic age of 35~kyr, the obtained size agrees well 
with the reported correlation  \citep{bamba2010}.

\subsection{Association with the supernova remnant G54.4$-$0.3}
\begin{figure}
\begin{minipage}[h]{0.49\textwidth}
\includegraphics[width=1.\linewidth, clip]{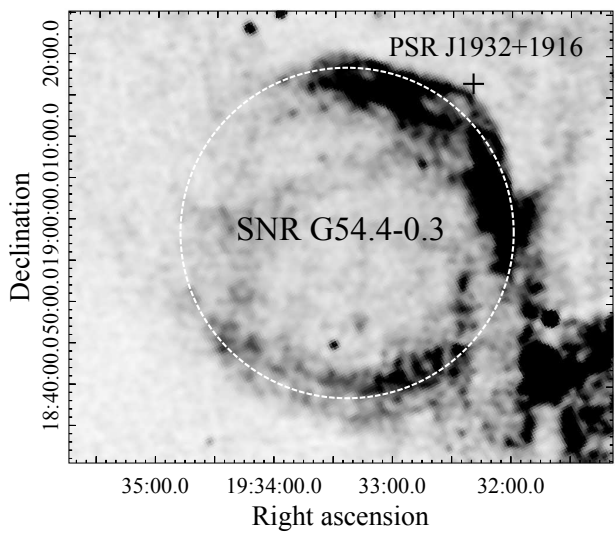}
\end{minipage}
\caption{Image of the  SNR G54.4$-$0.3 obtained from the
VLA Galactic Plane Survey (VGPS). 
The \textit{Fermi} position of \j1932
is marked by the cross. The dashed circle shows the SNR shape. 
}\label{fig:SNR}
\end{figure}

\j1932\ is located near the edge of  
the supernova remnant (SNR) G54.4$-$0.3. 
The radio image of the remnant  from the VLA Galactic Plane Survey
\citep{stil2006} is shown in Figure~\ref{fig:SNR}. 
In the radio, it is a shell-type SNR with a nearly
circular shape   and  angular
diameter of $\sim$ 40 arcmin \citep*{Junkes1992a}. 
Using  the \textit{ROSAT} observations of the remnant, \citet{junkes1996} estimated
the absorbing column density in its direction of about
$10^{22}$~cm$^{-2}$. 
This value is comparable to our $N_{\rm H}$ estimate for \j1932\
pointing to the similar distances and allowing us to assume that 
\j1932\ is associated with G54.4$-$0.3.

In this case, independent distance constraints are possible.
Based on  CO observations,  \citet{Junkes1992a} associated 
G54.4$-$0.3 with a molecular cloud at the velocity in the local
standard of rest (LSR) $v_{\rm LSR}=40$~km~s$^{-1}$. This led to the
kinematic distance of $3.0\pm 0.8$~kpc (near-side) or $7.0\pm
0.8$~kpc (far-side) using the Galaxy rotation curve of Burton
(1988). \citet{Park2013} revisited this value using rotational
curves of \citet{Brand1993} and \citet*{Levine2008} and obtained
$3.1$~kpc and $3.9$~kpc, respectively,  
for the near side of the rotational curve. 
Finally they adopted the distance of $3.3$~kpc given by \citet{Case1998}. 
The modern rotation curve
from \citet{Reid2009}\footnote{http://bessel.vlbi-astrometry.org/revised\underline{
\ }kd} gives $D=4.7\pm1.5$~kpc (near-side) and $D=5.0\pm 1.5$~kpc
(far-side).  Finally, the Bayesian distance estimator
\citep{Reid2016}\footnote{http://bessel.vlbi-astrometry.org/bayesian}, 
which assumes spiral arm associations, provides $D=4.0\pm0.4$~kpc
placing the SNR into the Local Spur.  
This distance is also supported by the presence of an OB association  
around G54.4$-$0.3 at the distances of
3$-$4~kpc \citep{Junkes1992a}. 

All this is compatible with our distance estimates for \j1932. 
Note, that it is natural to associate \j1932\ with the OB association. 
Even if the pulsar is not related to  G54.4$-$0.3, its progenitor star was likely 
the member of this association. 

Using the Inner-Galaxy Arecibo $L$-band Feed Array HI survey,
\citet{Park2013} found an expansion of the G54.4$-$0.3 shell. 
The measured expansion velocity allowed them to
estimate the dynamical age of the SNR $t_{\rm dyn}\approx
95$~kyr (for $D=3.3$~kpc). This  is by a factor of 2.7   larger 
than the characteristic age of  \j1932. One can consider this as a
counter-argument for the association,  
however there are examples where pulsar spin-down ages are considerably smaller than  
associated SNR ages \citep[e.g.,][]{lin2006}.   

The most natural way to check the reality of this association is
to measure the pulsar proper motion. In case of the association,
it should be  about 34$\tau_{35}^{-1}$ mas~yr$^{-1}$, where
$\tau_{35}$ is \j1932\ age in units of 35~kyr. 
Then the pulsar tangential velocity is $\approx 650 D_{4\rm kpc} \tau_{35}^{-1}$
km~s$^{-1}$, compatible with the empirical transverse velocity distribution for young pulsars \citep{Hobbs2005MNRAS}.
If the proper motion vector  points to the centre of
the remnant, it will be a strong argument in favour of the
association. Even if the proper motion is not detected, it will be 
possible  to estimate its direction basing on the PWN morphology.
For the PSR travelling sufficiently fast, one expects a bow-shock
morphology with a long trail \citep[e.g.,][]{gaensler2006}. 

\section{Conclusions}\label{S:Conclusions}

We analysed the archival \textit{Swift} and \textit{Suzaku} X-ray
observations of the  radio-quiet gamma-ray
pulsar \j1932$+$1916 field. At the pulsar position, \textit{Swift} found the
point-like X-ray source, while \textit{Suzaku} revealed diffuse emission with
extent of about 5 arcmin (FWHM). The spectral properties of these
sources are compatible with those of PSR$+$PWN systems suggesting 
the pulsar nature of the detected emission.

The analysis of the interstellar extinction allowed us to constrain
the distance to the pulsar within the range of $2-6$~kpc. 
At this distance, the observed
PWN candidate appears somewhat more luminous than PWNe powered by
pulsars with similar $\dot{E}$  while the candidate PSR
luminosity is in line with the values for other pulsars.

It is possible that \j1932\ is associated with the SNR
G54.4$-$0.3 located at $D\sim 3-4$~kpc. This association is
supported by  comparable values of the
interstellar extinctions towards \j1932\ and the SNR. 
If the presumable association 
is true, the prominent proper
motion of the pulsar can be expected. 

Future observations with better spatial
resolution are necessary to confirm \j1932\ counterpart nature of the
X-ray source by deeper studies of its morphology and spectral properties.

\section*{Acknowledgments}
We thank Dmitrii Barsukov, Serge
Balashev and Aida Kirichenko for discussions. The work was supported by the Russian
Science Foundation, grant 14-12-00316. 
For Figure~\ref{fig:SNR} we used the data from the VGPS survey conducted by the 
National Radio Astronomy Observatory (NRAO) 
instruments. NRAO is a facility of 
the National Science Foundation operated under cooperative
agreement by Associated Universities, Inc.


\bibliographystyle{mnras}
\bibliography{ref1932}

\clearpage

\end{document}